# Determination of layer-dependent exciton binding energies in few-layer black phosphorus

Guowei Zhang,[1,2] Andrey Chaves,[3] Shenyang Huang,[1,2] Fanjie Wang,[1,2] Qiaoxia Xing,[1,2] Tony Low,[4] Hugen Yan[1,2]*

[1]*Department of Physics, State Key Laboratory of Surface Physics and Key Laboratory of Micro and Nano Photonic Structures (Ministry of Education), Fudan University, Shanghai 200433, China*

[2]*Collaborative Innovation Center of Advanced Microstructures, Nanjing 210093, China*

[3]*Departamento de Física, Universidade Federal do Ceará, Caixa Postal 6030, Campus do Pici, 60455-900 Fortaleza, Ceará, Brazil*

[4]*Department of Electrical & Computer Engineering, University of Minnesota, Minneapolis, Minnesota 55455, USA*

*Corresponding author. Email: hgyan@fudan.edu.cn (H.Y.)

The attraction between electrons and holes in semiconductors forms excitons, which largely determine the optical properties of the hosting material, and hence the device performance, especially for low-dimensional systems. Mono- and few-layer black phosphorus (BP) are emerging two-dimensional (2D) semiconductors. Despite its fundamental importance and technological interest, experimental investigation of exciton physics has been rather limited. Here, we report the first systematic measurement of exciton binding energies in ultrahigh quality few-layer BP by infrared absorption spectroscopy, with layer (L) thickness ranging from 2-6 layers. Our experiments allow us to determine the exciton binding energy, decreasing from 213 meV (2L) to 106 meV (6L). The scaling behavior with layer number can be well described by an analytical model, which takes into account the nonlocal screening effect. Extrapolation to free-standing monolayer yields a large binding energy of ~800 meV. Our study provides insights into 2D excitons and their crossover from 2D to 3D, and demonstrates that few-layer BP is a promising high-quality optoelectronic material for potential infrared applications.

**Introduction**

Due to much stronger spatial confinement and reduced dielectric screening, excitons in atomically thin semiconductors are typically robust, showing binding energy one order of magnitude larger than its bulk counterpart. Moreover, the exciton energy spectrum deviates strongly from the two-dimensional (2D) hydrogen model (*1-3*). Of fundamental interest and importance is the question of dimensionality crossover for excitonic effects as the thickness of 2D semiconductor increases. However, such attempt for the widely studied 2D transition metal dichalcogenides (TMDCs), such as $MoS_2$, $WS_2$, $MoSe_2$ and $WSe_2$, is inherently challenging, due to the fact that the band gap is direct only for monolayers (*4*). Mono- and few-layer black phosphorus (BP) are emerging 2D direct-gap semiconductors (*5-8*), with the gap size strongly depending on the layer thickness (*9-12*). Therefore, BP provides us an ideal platform to interrogate 2D excitons and the crossover to 3D. Moreover, the anisotropy of 2D BP opens an avenue for investigation of anisotropic 2D excitons (*13, 14*), which are expected to show exotic sequences of excited energy levels. Although these unique features of excitons in few-layer BP are of great importance for both fundamental research and device applications, the exciton binding energy is still far from conclusive. For monolayer BP on $SiO_2$, the exciton binding energy was first experimentally reported to be as large as 0.9 eV (*15*), however, a more recent experiment indicates that the value is only 0.1 eV for monolayer BP covered by boron nitride (*10*). Moreover, the exciton binding energies in few-layer BP is not fully determined experimentally, though valuable information has been obtained from previous measurements by Li et al.(*10*).

In this work, we investigate the excitonic effects in few-layer BP on polydimethylsiloxane (PDMS) substrates with layer (L) number $N$ = 2-6, using infrared (IR) absorption spectroscopy (see Materials and Methods). We find that excitonic resonances dominate the optical absorption. For the high-quality BP samples studied here, the optical absorption spectrum exhibits surprisingly sharp and intense features, with an unprecedented linewidth of < 25 meV at room temperature for the band edge excitons. Such high optical quality was previously unattainable, due to the degradation of few-layer BP in ambient conditions (*16*). Most strikingly, the spectral features of lowest-energy excited states are clearly observable even at room temperature. In conjunction with numerical calculations, exciton binding energies are determined for 2-6L BP, and the binding energy for monolayer BP is also inferred. The scaling behavior of the binding energy with layer number is carefully examined. Our results pave the way for BP applications in infrared optoelectronics, such as photo-detection (*17-19*), optical modulation (*20*), lasing (*21, 22*) and polariton condensation (*23*).

## Results

**Exciton-dominated optical absorption**

In 2D semiconductors, the optical absorption is characteristic of step-like features in the single-particle picture (*24, 25*), with the onset of band-to-band transitions as the quasiparticle bandgap $E_g$. Electron-hole (e-h) interactions are found to dramatically modify the optical response, leading to new spectral features below $E_g$. Fig. 1B illustrates the optical absorption in a model 2D semiconductor, including transitions to exciton bound states and free carrier states (continuum), with the exciton binding energy defined as $E_b = E_g - E_{opt}$, where $E_{opt}$ is the optical gap, related to the optical transition energy of the ground state exciton.

Few-layer BP samples were directly exfoliated on PDMS substrates from bulk crystals (HQ Graphene, Inc.), with areas typically over 1500 $\mu m^2$, large enough for us to obtain accurate IR extinction ($1-T/T_0$) spectrum, where $T$ and $T_0$ denotes the light transmittance of samples on PDMS and bare PMDS, respectively. For atomically thin layers supported by a thick transparent substrate, when the optical conductivity is not large, the extinction ($1-T/T_0$) is approximately proportional to the real part of the optical conductivity (*26*). The layer number of BP flakes is determined by IR characterization (*9*). In order to obtain high optical quality of BP samples, we performed IR measurements right after the exfoliation, using a Fourier transform infrared spectrometer (FTIR) in combination with an IR microscope under ambient conditions. The exposure time in air with low humidity is typically < 5 minutes. Comparative studies for samples without exposure to air have been performed. We found that such short time exposure (< 5 minutes) to low humidity air has little effect on the optical quality. In our previous studies (*9*), BP layers were first exfoliated on PDMS and then transferred to other substrates. The quality of such samples were typically much lower.

A representative IR extinction spectrum of a 4L BP sample on PDMS is shown in Fig. 1C, with normal light incidence and polarization along the two characteristic directions (more data for 4L BP are presented in fig. S1). In few-layer BP, conduction and valence bands split into quantized subbands, due to quantum confinement in the out of plane direction, similar to the traditional quantum wells (QWs). Optical transitions with subband index difference $\Delta n$ = 0 for valence and conduction bands are allowed in symmetric QWs (*27, 28*). We use the symbol $E_{nn}$ denoting the optical transition $v_n \rightarrow c_n$ at the Γ point of the 2D Brillouin zone, illustrated in Fig. 2H. Strikingly, we observed a very sharp peak (labeled as $E_{11}$) in the extinction spectrum for armchair (AC) light polarization, with a linewidth as narrow as ~20 meV at room temperature. The $E_{11}$ peak has a Lorentzian lineshape, characteristic of exciton absorption (*2, 3, 29*). This is the ground state exciton transition. For clarity, we label the ground and excited states of excitons in analogy to hydrogenic Rydberg series as 1*s*, 2*s*, etc. The extinction at the $E_{11}$ resonance reaches ~12%, suggesting very strong light-matter interactions in such atomically thin BP layers. In addition to the 1*s* peak, a weak and broad peak can be clearly resolved on the high energy side, which we attribute to the 2*s* transition. Moreover, we can observe a relatively flat "plateau" above ~0.8 eV, this feature is attributed to the continuum band-to-band transitions (*30, 31*). It should be noted that the narrow spectral linewidth of the 1*s* state is the key to clearly identify a 2*s* state in the absorption spectrum, otherwise they will be merged. For zigzag (ZZ) light polarization, the spectrum is featureless as expected, with a slightly tilted non-resonant background. Owing to the low symmetry of the crystal structure, the excitonic response of BP is strongly polarization dependent (*9, 10, 12, 15*) (see fig. S2).

In order to probe the layer dependence of excitonic response, we performed IR absorption measurements on few-layer BP from 2L to 6L, as shown in Fig. 2 (A-E). The band edge optical transitions in all samples exhibit very narrow spectral features, indicating very high sample quality. The $E_{11}$ resonances are all spectrally sharp and intense. Most importantly, the 2*s* features can be clearly identified in the IR spectra as indicated by red arrows. Thus, the 1*s*-2*s* separation ($\Delta_{12}$) can be directly extracted and the value is 122, 87, 76, 62 and 58 meV for 2-6L samples, respectively. As demonstrated previously (*9, 10*), the optical transition energies exhibit remarkable layer dependence, owing to the strong interlayer interaction (*32*). For comparison, we also performed similar IR measurements for thicker samples (7L and 8L), as shown in Fig. 2 (F and G). The 1*s* peaks are as sharp, but the 2*s* exciton features are unresolvable, presumably due to a smaller 1s-2s energy separation. Moreover, for $E_{22}$ transitions, the exciton peak is broader, indicating a shorter exciton lifetime for higher energy transitions. This is reasonable since carriers at the higher subband would energetically relax towards the band edges. Nevertheless, the $E_{22}$ peaks are still very clear, in sharp contrast to the featureless band-to-band continuum transitions, a manifestation of exciton transitions as well.

**Extraction of the exciton binding energy**

In a 2D hydrogen model, it's well known that the 1$s$-2$s$ separation accounts for 8/9 of the total exciton binding energy. With the measured 1$s$ and 2$s$ transition energies, the exciton binding energy would then become $E_b$ = 9/8$\Delta_{12}$. However, this picture was found to break down in atomically thin 2D crystals, such as TMDC monolayers, due to the nonlocal dielectric screening of Coulomb interactions (*1-3*). In 2D case, both the material itself and its surroundings contribute to the dielectric screening. Generally, the dielectric constant of the material is far greater than that of its surroundings. For higher-order excited excitonic states, the e-h spatial separation is larger, so the electric field lines would experience a larger portion of screening from the surroundings. Consequently, the effective screening is significantly reduced, and this effect is also known as the anti-screening effect (*1*). The non-uniform dielectric environments result in a pronounced deviation of the excitonic states from the 2D hydrogenic Rydberg series and the simple relation $E_b$ = 9/8$\Delta_{12}$ doesn't hold. Nevertheless, the observed 2$s$ features in our experiments are still very informative, since the energy separation from the 1$s$ transition can give us a lower bound of the exciton binding energy. In fact, in combination with numerical calculation, we are able to determine the exciton binding energies, and hence the quasiparticle bandgaps.

In order to extract the exciton binding energy, we calculate the excitonic states in few-layer BP on PDMS substrates ($N$ = 2-6), within the Wannier-Mott framework (see the Supplementary Materials for details), by solving the Schrödinger equation numerically (*14, 33*). As shown in Fig. 3A, the calculated 1*s*-2*s* separation agrees very well with our experimental observations, confirming the assignment of the weak features to 2*s* transitions. Furthermore, the exciton binding energies are extracted to be 213, 167, 139, 120 and 106 meV, respectively. Even larger values are expected for suspended samples, since generally $E_b$ is inversely proportional to $\varepsilon^2$ ($\varepsilon$ is the dielectric constant) and the screening from the underlying substrate is expected to significantly reduce the exciton binding energy. Such large binding energy in few-layer BP is one order of magnitude larger than that in bulk BP (*34*), mainly originating from the weak dielectric screening and strong quantum confinement in 2D case. The binding energies determined here are in good agreement with previous theoretical calculations (*14, 35*).

Interestingly, one may notice that the 1*s*-2*s* separation only accounts for about 1/2 of the total exciton binding energy, rather than 8/9 predicted in 2D hydrogen model. In other words, the observed exciton series in few-layer BP is non-hydrogenic, a very general feature in 2D semiconductors (*3*). In addition, the exciton binding energy decreases with increasing layer number, showing strong thickness dependence. Relative to the 1s transition energy (optical gap), we extracted the quasiparticle bandgap for 2-6L BP as $E_g = E_{opt} + E_b$, the values are 1.33, 1.01, 0.84, 0.73 and 0.66 eV, respectively, summarized in Fig. 3B, exhibiting strong layer dependence as predicted by previous calculations (*11, 12*).

**Layer dependence of the exciton binding energy**

The layer dependence of the exciton binding energy is of particular interest, given that it can shed light on the dimensional crossover from 2D to 3D for excitons. Recently, Olsen et al. proposed a simple screened hydrogen model to describe excitons in 2D materials (*36*). $E_b$ has an analytical expression as a function of the reduced effective mass $\mu$ of excitons and the sheet polarizality $\alpha$: $E_b = \dfrac{2\mu}{(m-1/2)^2(1+\sqrt{1+\dfrac{32\pi\alpha\mu}{9m(m-1)+3}})^2}$, where $m$ is the s-state index and the equation is in atomic units. Under the condition $32\pi\mu\alpha/3 \gg 1$, $E_b$ for ground state exciton ($m = 1$) can be simplified as $E_b \approx \dfrac{3}{4\pi\alpha}$, which is proved to be valid over a wide range of 2D TMDCs (*36*) and should also work for few-layer BP. This indicates that the exciton binding energy of 2D materials is solely determined by its polarizability, and does not directly depends on the effective mass, in sharp contrast to the 3D case. The 2D polarizability relates to the screening effect and can be viewed as the in-plane component of the 3D polarizability in its bulk counterparts. The effect of the substrate can also be included, by simply replacing the polarizability $\alpha$ with an effective value $\alpha_{eff}$. This effective polarizability should also depend on the substrate, in addition to the polarizability of the 2D material. For a few-layer BP, we can still treat it as a 2D system in the limit where the thickness is small compared to the exciton Bohr radius. Simple estimation based on the effective masses and dielectric constant of bulk BP gives a Bohr radius of ~5 nm (*12, 37*), which validates the approximation in our case. Since the polarizability of BP is approximately equal to its density of states (*38*), it scales roughly with the layer number. Hence, we propose a layer-dependent effective sheet polarizability $\alpha_{eff}(N) = \alpha_0 + N\alpha_1$, where $\alpha_0$ and $N\alpha_1$ accounts for the screening from the underlying substrate and the 2D material itself, respectively. With this linear relation between $\alpha_{eff}$ and $N$, we have (in atomic units)

$$E_b = \frac{3}{4\pi(\alpha_0 + N\alpha_1)} \tag{1}$$

Fig. 4 plots numerically obtained $E_b$ as a function of $N$. In addition to the data points shown in Fig. 3, more data points for 1L, 7L and 8L samples are also shown here, obtained using the same numerical methods (see the Supplementary Materials). We use Eq. 1 to fit our data. With fitting parameters $\alpha_0$ = 6.5 Å and $\alpha_1$ = 4.5 Å, it shows excellent agreement. The value for $\alpha_1$ extracted here matches well with previous calculations ($\alpha_1$ = 4.1 Å) (*13*). These parameters are very reasonable given that the bulk dielectric constant $\varepsilon$ for BP is ~$10\varepsilon_0$ (*37*). Compared to the 3D counterparts, mainly two factors determine the exciton binding energies for a 2D material, namely, the reduced screening and spatial confinement in the out-of-plane direction. When increasing the layer number, Eq. 1 doesn't take into account the reduced confinement and only the increase of the screening is considered. The good agreement with data shows that in the small thickness limit, the screening effect plays a dominant role. Of course, the model breaks down when the thickness is much larger than the Bohr radius and a 3D-type exciton model will be required to further describe the crossover from 2D to 3D.

With the fitting parameter $\alpha_1$ = 4.5 Å and by setting $\alpha_0$ = 0, we can infer the binding energies for free-standing few-layer BP: $E_b$ = 762/$N$ meV, as shown in fig. S3. The free-standing monolayer BP has an exciton binding energy of 762 meV, in good agreement with ab initio calculations based on the Bethe-Saltpeter equation framework, which gives $E_b$ ~ 800 meV (*12*). $E_b$ for monolayer BP on PDMS is 316 meV, less than half of that for free-standing samples, indicating the importance of substrate screening effect. Our systematic study does not agree with the previous study of excitons in monolayer BP on $SiO_2$ (*15*), which claims a much larger $E_b$ of 900 meV, despite similar dielectric constants of $SiO_2$ and PDMS. Certainly, further experiments would help in resolving this discrepancy.

## Discussion

In summary, with ultrahigh optical quality samples, we successfully determined the exciton binding energies for 2-6L BP. The large binding energy we observed here shows strong layer dependence, in good agreement with a model which solely takes into account the linear increase of the sheet polarizability in the quasi-2D limit. This sheds light on the exciton crossover from 2D to 3D. The strong excitonic effects are expected to profoundly impact the device performance in BP optoelectronics even at room temperature. Few-layer BP provides us an ideal 2D system to study the layer dependent many-body physics, such as excitons and trions (charged excitons). Moreover, as a high optical quality anisotropic 2D system, few-layer BP opens an avenue to study anisotropic 2D excitons (*13, 14*), where light polarization can be leveraged to selectively excite excitons with different angular momenta. The ultrahigh sample quality facilitates us to probe the dark excitonic states in few-layer BP using two-photon photoluminescence excitation spectroscopy (*2, 3, 29*).

## Materials and Methods

### Polarized IR absorption spectroscopy

The polarization-resolved IR measurements on few-layer BP were performed using a Bruker FTIR spectrometer (Vertex 70v) equipped with a Hyperion 2000 microscope. A tungsten halogen lamp was used as the light source to cover the broad spectral range 3750–11000 cm$^{-1}$ (0.36–1.36 eV), in combination with a liquid nitrogen cooled MCT detector. The lower bound cutoff photon energy is restricted by the PDMS substrate. The incident light was focused on BP flakes using a 15X IR objective, with the polarization controlled by a broadband ZnSe grid polarizer. All the measurements were conducted at room temperature under ambient conditions with low humidity (< 30%).

## Supplementary Materials

- note S1. Anisotropic optical absorption.
- note S2. Numerical calculation of the exciton binding energy.
- note S3. Substrate screening effect on the exciton binding energy.
- fig. S1. Additional IR extinction data for 4L and 6L BP.
- fig. S2. Polarized IR extinction spectra for an 8L BP.
- fig. S3. Comparison of exciton binding energies for suspended and supported few-layer BP.

- table S1. Summary of experimentally and theoretically obtained values for 1$s$-2$s$ separation ($\Delta_{12}$), as well as the exciton binding energy ($E_b$).

- References (*39–43*)

## Acknowledgments

**Funding:**

H. Y. is grateful to the financial support from the National Young 1000 Talents Plan, the National Key Research and Development Program of China (Grant numbers: 2016YFA0203900, 2017YFA0303504), and Oriental Scholar Program from Shanghai Municipal Education Commission. G. Z. acknowledges financial support from China Postdoctoral Science Foundation (Grant number: 2016M601489). A. C. acknowledges financial support from CNPq, through the PQ and PRONEX/FUNCAP programs. Part of the experimental work was carried out in Fudan Nanofabrication Lab.

**Author contributions:**

H. Y. initiated the project and conceived the experiments. G. Z. prepared the samples and performed the measurements with assistance from S. H., F. W. and Q. X.. T. L. and A. C. provided modeling and theoretical support. H. Y., G. Z., T. L. and A. C. performed data analysis. H. Y. and G. Z. co-wrote the manuscript with inputs from T. L. and A. C.. H. Y. supervised the whole project.

All authors commented on the manuscript.

# Figures and figure captions

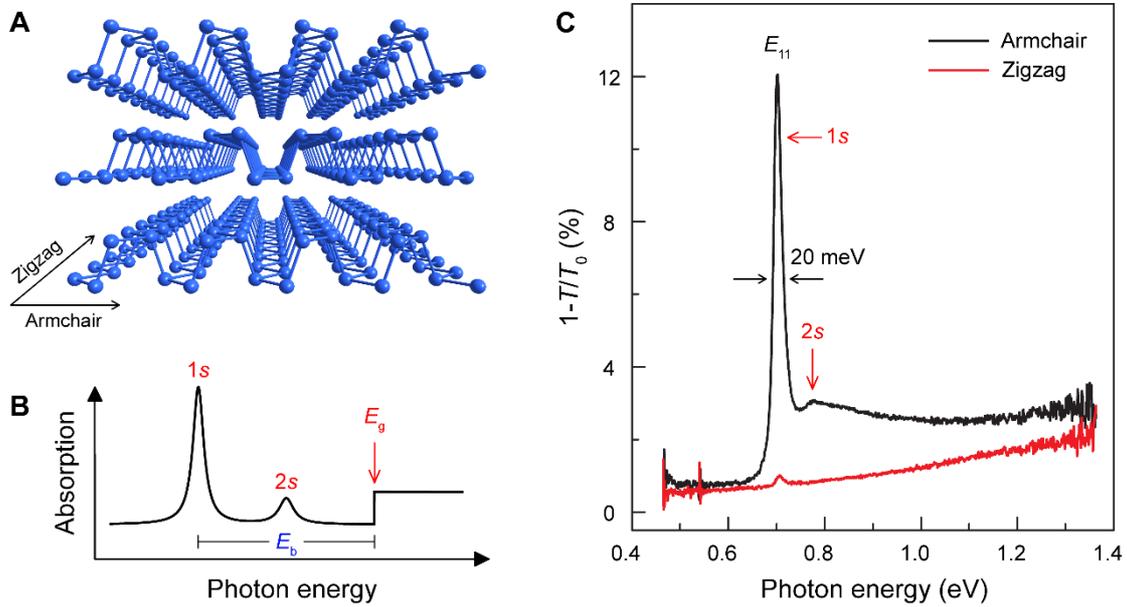

**Fig. 1. Optical absorption in 2D semiconductors.** (**A**) Lattice structure of few-layer BP, showing the puckered hexagonal crystal with two characteristic directions: armchair (AC) and zigzag (ZZ). (**B**) Illustration of optical absorption in a model conventional 2D semiconductor, including optical transitions to excitonic ground (1s) and excited states (2s), as well as continuum states above the quasiparticle bandgap $E_g$. The exciton resonance is characteristic of a Lorentzian lineshape, while the continuum absorption exhibits a step-like feature. (**C**) IR extinction spectra (1-$T/T_0$) for a representative 4L BP sample on a PDMS substrate, with two different light polarizations. The symbol $E_{11}$ denotes exciton relating to optical transition $v_1 \rightarrow c_1$, as illustrated in Fig. 2H. Data were collected at room temperature.

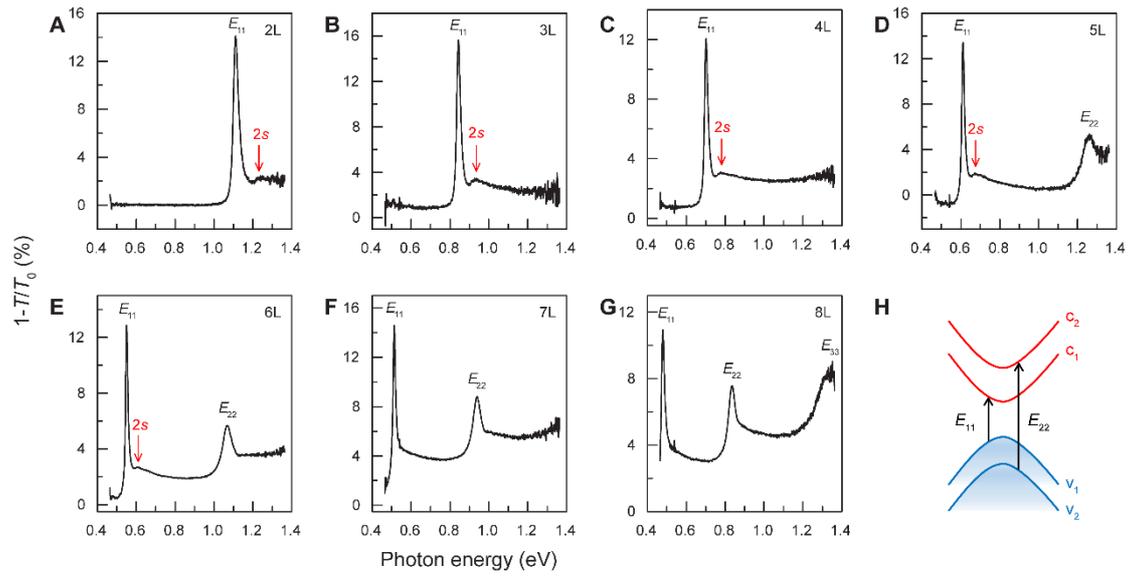

**Fig. 2. Layer-dependent IR extinction spectra.** (**A–G**) IR extinction spectra ($1-T/T_0$) for few-layer BP on PDMS substrates with layer number $N$ = 2-8, with incident light polarized along the AC direction. The red arrows indicate the lowest-energy excited states (2s) of the $E_{11}$ transition. For 7L and 8L samples, the 2s peaks are unresolvable. (**H**) Schematic illustration of optical transitions between quantized subbands in few-layer BP, with the symbols $E_{11}$ and $E_{22}$ denoting transitions $v_1 \rightarrow c_1$ and $v_2 \rightarrow c_2$, respectively.

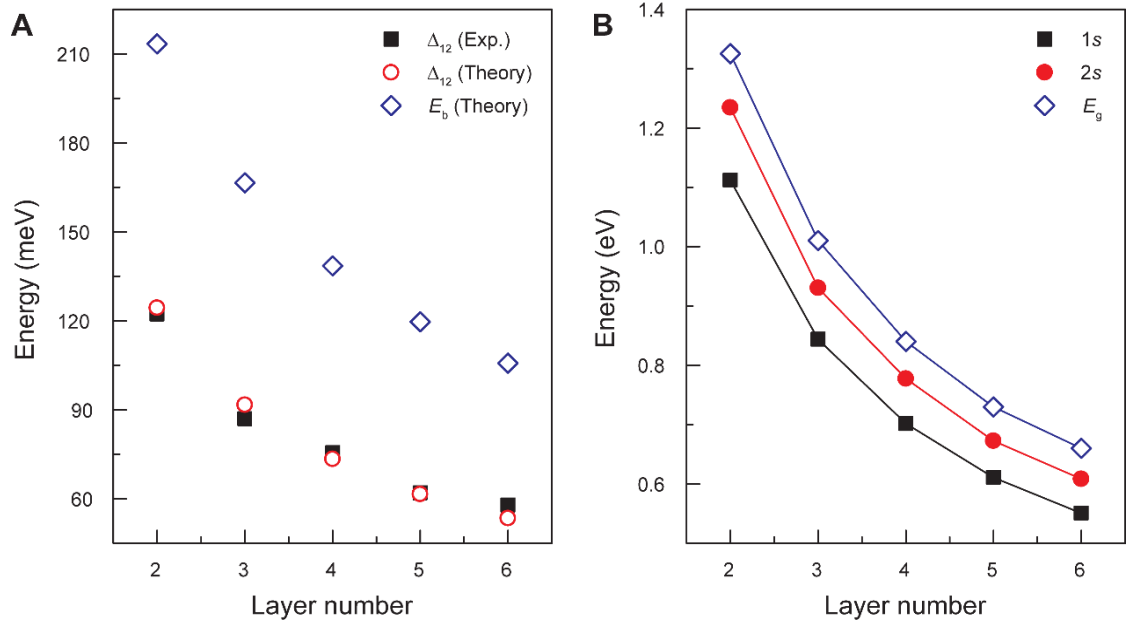

**Fig. 3. Extraction of the exciton binding energies and quasiparticle bandgaps.** (**A**) Experimentally and theoretically obtained 1s-2s separation ($\Delta_{12}$) as a function of layer number. The blue squares denote the exciton binding energy obtained using a numerical method detailed in the Supplementary Materials. (**B**) Layer dependence of the 1s and 2s transition energies and the quasiparticle bandgap. The 1s and 2s transition energies are obtained experimentally, as indicated in Fig. 2. The quasiparticle bandgap is deduced as $E_g = E_{opt} + E_b$, where $E_{opt}$ is the 1s transition energy.

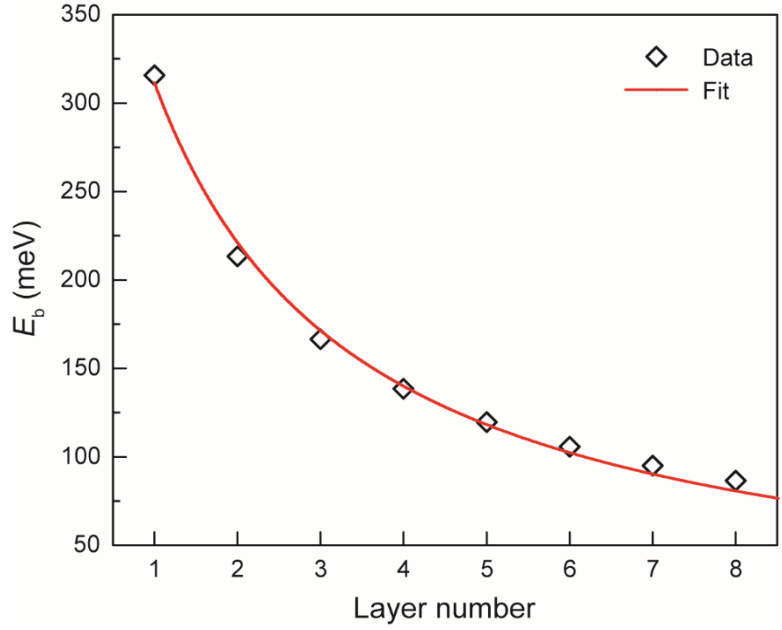

**Fig. 4. Scaling behavior of the exciton binding energy with layer number.** The black squares are theoretical values for few-layer BP with $N$ = 1-8, obtained within the Wannier-Mott framework. The red line is a fit to the data, using the formula $E_b = 3/4\pi\alpha_{eff}$. $\alpha_{eff}$ is the effective 2D polarizability, with a linear function of layer number $N$: $\alpha_{eff} = \alpha_0 + N\alpha_1$, where $\alpha_0$ and $N\alpha_1$ describes the screening effect from the underlying substrate and the 2D material itself, respectively.